\newcommand{\R}{\overline{R}}
\newcommand{\bm}{\begin{displaymath}}
\renewcommand{\em}{\end{displaymath}}
\newcommand{\Lambd}{{\mit\Lambda}}
\newcommand{\lora}{\longrightarrow}
\newcommand{\ot}{\otimes}
\documentstyle[11pt]{article}
\begin{document}
\begin{flushright}\footnotesize{LMU-TPW 93-28}\end{flushright}
\begin{center}{\bf
POINCAR\'{E} SERIES OF QUANTUM MATRIX BIALGEBRAS DETERMINED BY A PAIR
OF QUANTUM SPACES}\\
\end{center}\vspace{6ex}
\begin{center}
{\bf Phung Ho Hai}\\
Section Physik der Universit\"{a}t M\"{u}nchen\\
Theresienstr 37, 80333 M\"{u}nchen
\\ {\small e-mail: phung@mathematik.uni-muenchen.d400.de}
\date{28.02.1994}
\end{center}\vspace{2ex}

\newtheorem{dl}{Theorem}[section]
\newtheorem{cor}[dl]{Corollary}
\newtheorem{pro}[dl]{Proposition}
\newtheorem{lem}[dl]{Lemma}

\begin{abstract}{\small
  The dimension of the third  homogeneous  component  of a  matrix
  quantum bialgebra, determined  by  pair of  quantum spaces,
  is calculated. The Poincar\'{e} series of some deformations
  of $GL(n)$ is  calculated. A new deformation of $GL(3)$ with the correct
dimension is given.}
\end{abstract}

\section{Introduction}
\label{section-intro}
  Quantum groups are the subcategory of the category of Hopf-algebras,
which are generally
non-commutative and non-cocommutative and possess many properties similar to
those of classical universal enveloping algebras or, by duality, function
rings of affine algebraic groups. There are many approaches to quantum groups,
the "R-matrix" and Quantum
Yang-Baxter equations approach, which was introduced by Faddeev, Takhtajan,
Reshetikhin and others, formal deformations of commutative or cocommutative
Hopf algebras especially of the universal enveloping algebra $U(g)$ of a Lie
algebra $ g$, which was systemmatically studied by
Drinfeld, Jimbo and others. Quantum groups can be also considered as the
symmetries of quadratic
non-commutative or quantum spaces. This construction was given by Yu.Manin
[M2].
The methods in this paper are based on Manin's construction.

 A family of quantum spaces defines an algebra, which is universally
coacting upon  the spaces of this family. This algebra because of its
universality is
a bialgebra. We call it a quantum matrix bialgebra (QMB).
 To obtain a Hopf algebra envelope
of this bialgebra, one often has to invert a certain
 quantum determinant, which is
a group like element and in many examples a central element. In this paper we
 restrict ourselves to an investigation of the quantum
matrix bialgebra, which induces deformations of $GL(n)$. More concretely we
study a quantum matrix bialgebra, which is
determined by two quantum spaces, which generalize a pair of symmetric and
antisymmetric tensor algebras on a linear space. It is natural to consider the
above QMB as a deformation of the bialgebra $M(n)$ (a polynomial ring on the
semigroup of linear transformations of a vectorspace). Since our bialgebra
is a graded algebra, the question about its Poincar\'e series arises.

Note that in our construction we obtain an operator -- "R-matrix".
We always require it to
satisfy the so called quantum
Yang-Baxter equation. This operator plays an important
r\^ole in quantum theory. For deformations of $M(n)$ it should satisfy
the Hecke equation. In the undeformed case, for $M(n)$, it is just the
twist operator.

 Recently several authors have considered the polynomiality and the
Poincar\'{e} series of a quantum matrix bialgebra by using the diamond lemma.
 The
troubles usually happen on cubic monomials. In general, the method of the
diamond lemma seems not to be effective because there is no sense of
reduction system. In this work we do the first step of calculating
the Poincar\'e series without using the diamond lemma. We use the method
developed by Sudbery[S2]. We get a formula for calculating the
dimension of the third component of the QMB via the dimension of the
 components
of the quantum spaces (Section 3).
Nevertherless, we remark that
the obtained formula is independent of the matrix R.
It seems to be complicated to do the same
for the higher components of the QMB. It should be interesting
to show that the same independence takes place for all  higher
homogeneous components. If this would be true, we could say
that if the quantum spaces have the correct Poincar\'e series
then the QMB has the correct Poincar\'e series, too.

On the other side, knowing the dimension of the second and third components
 in some cases we can already obtain the Poincar\'e series of the
QMB. Applying Bondal's theorem ([Bo], Theorem 3) to $M(n)$ we have, if
$R=R(q), q\in {\bf R}$ is some continuously parameterized deformation of $P$ --
the twist operator, $R(1)=P$, and if the Poincar\'e series of
the quantum spaces are correct, then for every interger n we can choose a
neighborhood of 1, such that the dimensions of the homogeneous components
of the QMB obtained from $R(q)$
with $q$ in this neighborhood is correct up to the n-th component.

 In section 4 we investigate the polynomiality of all known QMBs which give
a deformation of $GL(n)$. The polynomiality of the multiparameter
deformations was shown in [ATS]. We show here the
polynomiality of a deformation obtained by Creme and Gervais [CG]. At the end
of the section we give a new deformation of $GL(3)$.

\section{A Yang-Baxter operator on a universal bialgebra.}
\label{sec-manin}\par
 Let $k$ be a fixed algebraically closed
field, $V$ be a $k$-vector space of finite dimension $n$.
A quadratic algebra $A$ is a factor algebra
of the tensor algebra $T(V)$ by a two-sided ideal generated
by a subspace $R_{A}$ of $V\otimes V$. Hence $A$ is a graded algebra:
\begin{displaymath}
A= \bigoplus_{i=0}^{\infty}A_{i},
A_{0}=k, A_{1}=V
\end{displaymath}
We call the pair $(A,V)$ a quantum space.

Let $B$ be a k-algebra. A left coaction of $B$ on the quantum space $(A,V)$
is an algebra homomorphism \[ \delta_{B}:A \longrightarrow B \otimes A \]
such that
\[ \delta_{B}(V)\subset B \otimes V \]
A coaction of $B$ on a family of quantum spaces
 $(A_{\alpha},V), \alpha \in J$
is a family of coactions of $B$ on each $(A_{\alpha},V), \alpha \in J$,
such that their restrictions on $V$ coincide.

An algebra $E$ is said to universally coact upon
$(A_{\alpha},V)$ if for every algebra $B$ coacting
on $(A_{\alpha},V), \alpha \in J$,
there exists a unique algebra homomorphism
$\phi :E\rightarrow B$ such that
$(\phi\otimes id)\delta_\alpha =\delta^{'}_\alpha, \alpha\in J$,
where $\delta_\alpha$, $\delta_\alpha^{'}$ denote the coactions
of $E$ and $B$  respectively.
 Mukhin [Mu] and Sudbery [S1] show that $E$ exists.
The bialgebra structure on $E$ follows from the universality.

Let  $\theta$ denote the isomorphism
$$\theta : V^{*}\otimes V^{*}\otimes V\otimes V\longrightarrow (V^{*}\otimes V)
^{\otimes 2} \cong \mbox{End}(V)\otimes \mbox{End}(V)$$
which interchanges the  2nd and 3rd components. Analogously we have the
isomorphism of the two spaces $(V^{*})^{\otimes 3}\otimes V^{\otimes 3}$
and  $\mbox{End}(V)^{\otimes 3}$, which is denoted by the same $\theta$.

In this paper we study certain deformations of $M(n)$. In this case there are
only two quantum spaces $(\Lambd ,V)$ and $(S,V)$ with
$R_{\Lambd}\oplus R_S =V^{\otimes 2}$.
$(\Lambd ,V)$ and $(S,V)$ play the analogous roles to the antisymmetric and
symmetric tensor algebras on the space $V$.
Let $E$ be the universal matrix bialgebra coacting on
$(\Lambd ,V)$ and $(S,V)$. Then $E$ is a quadratic algebra over $E_1=V^*\ot V.$
The relations on $E$ are given by $R_E=\theta(R_S^\perp\otimes R_S+
R_{\Lambd}^\perp\otimes R_{\Lambd} )$.

 Let $P_{\Lambd}$ and $P_{\mbox{s}}$ be
the projections onto $R_{\Lambd}$ and
$R_S$ respectively with respect to $R_{\Lambd}\oplus R_S =V^{\otimes 2}$.
Let  \bm
R=q P_S+P_{\Lambd}, \ q\neq 1.
\label{eq:6}
\em
And $R^*:V^*\ot V^*\lora V^*\ot V^*$.
$R$ and $R^*$ are plainly diagonalizable. Let $t_i, i=1,\ldots ,n^2$ be the
eigenvectors of $R$ which form a basis on $V\ot V$, the dual basis to it
consists of eigenvectors of $R^*$, denote them by $u_i, i=1,\ldots ,n^2$.
The vectors $u_i\ot t_j$ are eigenvectors of $R^*\ot R^{-1}$
and form a basis on $V^*\ot V^*\ot V\ot V$. Consequently
$$R_S^{\perp }\otimes R_S+R_{\Lambd}^{\perp }\otimes R_{\Lambd}=
\mbox{Im }(R^*\otimes R^{-1}-1 ).$$
Thus
\begin{equation} R_E=\theta(\mbox{Im}\sum_\alpha S_\alpha)=
\mbox{Im }(\theta(R^*\otimes R^{-1})\theta^{-1}
 -1). \label{eq:hai}\end{equation}
 We assume that our construction is Yang-Baxter, in other words,
 there exists a $q$ such that $R$ satisfies
 \begin{equation}\label{eq:YB}
R_{12}R_{23}R_{12}=R_{23}R_{12}R_{23},
   (R_{12}=R\otimes 1, R_{23}=1\otimes R).
  \end{equation} $R$ is called a Yang-Baxter operator.
And we assume that $q$ obeys the following condition
\begin{equation}
q(q-1)(q^{2}-q+1) \not=  0.
\label{eq:cond}
\end{equation}
For a quadratic algebra $A$ we have $A=\oplus_{i=0}^\infty A_i$.
Since the $A_i$ are all finite dimensional over $k$ we can
form a formal series
$P_A(t)=\sum_{i=0}^\infty (\mbox{dim}_kA_i)t^i$, which is called the
Poincar\'e series of $A$.\\
{\bf Definition.}

a) A pair of quadratic algebras $(S,V)$ and $(\Lambd ,V)$ is said
to be {\it compatible}
if their Poincar\'{e} series coincide with the ones of the symmetric
and antisymmetric tensor algebras over $V$ respectively.

b) The bialgebra E is said to be {\it compatible} if its Poincar\'{e} series
coincides with the one of the polynomial ring of $n^{2}$ parameters.

\section{Yang-Baxter operator with Hecke equation. The main theorem}
The aim of this section is to calculate the dimension of the third
homogeneous component
of $E$ via the dimensions of components of $\Lambd$ and $S$. Thus, let
$s_{i}  = \mbox{dim }S_{i},\
\lambda _{i}  =  \mbox{dim }\Lambd _{i}, i=2,3$
be given. We want to find $e_{3}  =  \mbox{dim }E_{3}$.

Let $R=qP_S+P_{\Lambd}$ be Yang-Baxter and let $q$ satisfy (\ref{eq:cond}).
Then $(R-q)(R-1)=0$.
\renewcommand{\L}{\rule{.8ex}{.02ex}\hspace{-.8ex}\rule{.02ex}{.8ex}
\hspace{.8ex}}
\renewcommand{\Gamma}{\rule[.8ex]{.8ex}{.02ex}\hspace{-.8ex}\rule{.02ex}{.8ex}
\hspace{.8ex}}
Consider the following operators
\bm P_{\alpha}=P_{\alpha}(R)=(R_{12}-\alpha )
(R_{23}R_{12}-\alpha R_{23}+\alpha ^{2}) (\alpha =1,q)
\label{eq:palpha}
\em
\bm P_{\Gamma}=P_{\Gamma}(R)  =  (R_{12}-q)(R_{23}-1)
\label{eq:pg}
\em
\bm P_{\L} =P_{\L}(R) =  (R_{23}-q)(R_{12}-1)
\label{eq:pl}
\em
Then $R_{13}P_{\L}=P_{\Gamma}R_{13}$,
where $R_{13}=R_{12}R_{23}R_{12}=R_{23}R_{12}R_{23}.$
By $\Pi _{\alpha}$ we denote the subspaces $ \mbox{Im} P_{\alpha}$
of $V^{\otimes 3}, (\alpha =1,q,\L ,\Gamma )$. The notations $\L$ and $\Gamma$
are due to Sudbery, they are chosen for their resemblance to the Young tableau.

 Analogously we define $P_{\alpha}^{+}=P_{\alpha}(R^{*})$ and
$\Pi _{\alpha}^{+}=\mbox{Im} P_{\alpha}^{+}$, $\alpha =1,q,\L ,\Gamma$.
$R^*$ is the adjoint operator of $R$.

For $\alpha =1,q$ we have
\begin{equation}
\begin{array}{rl}
P_{\alpha}=&(R_{12}-\alpha )(R_{23}R_{12}-\alpha R_{23}+\alpha ^{2})\\
 =&(R_{23}-\alpha )(R_{12}R_{23}-\alpha R_{12}+\alpha ^{2})\\
 =&(R_{23}R_{12}-\alpha R_{12}+\alpha ^{2})(R_{23}-\alpha )\\
 =&(R_{12}R_{23}-\alpha R_{23}+\alpha ^{2})(R_{12}-\alpha )
\end{array}\label{eq:8}
\end{equation}
The following lemma is due to Sudbery.
\begin{lem}  ([S2], Lemma 2)
\label{sec-sudbery}
There exist constants $c_{\alpha}$ such that
$c_{\alpha}P_{\alpha}$ are projections onto $\Pi _{\alpha}$ and
$\oplus \Pi _{\alpha}=V^{\otimes 3},
\ \alpha =1,q,\Gamma , \L .$
\end{lem}

 From the definition we have
\begin{equation} s_2=\mbox{dim Im }(R-q), \lambda _2=\mbox{dim Im }(R-1).
\label{eq10}
\end{equation}
We claim
\bm
\mbox{Ker} P_{q}=\mbox{Im }(R_{12}-1)+ \mbox{Im }(R_{23}-1).
\em
Indeed if $x\in \mbox{Ker } P_{q}:
(R_{12}-q)(R_{23}R_{12}-qR_{23}+q^{2})x=0$, then
\begin{eqnarray*}
 (R_{12}-1)(R_{23}R_{12}-qR_{23}+q^{2}-q+1)x-(1-q)
 (R_{23}-1)(R_{12}-q)x=\\ (q-1)(q^{2}-q+1)x .\end{eqnarray*}
Hence, if $(1-q)(q^{2}-q+1)\not= 0$,
 $x \in \mbox{Im }(R_{12}-1)+ \mbox{Im }(R_{23}-1).$
Conversely, if $x \in \mbox{Im }(R_{12}-1)+ \mbox{Im }(R_{23}-1),$
then $x=x'+x'',$ where
$x' \in \mbox{Im }(R_{12}-1),x''\in  \mbox{Im }(R_{23}-1).$
And obviously $P_{q}x=P_{q}x'+P_{q}x''=0.$
Consequently
\begin{equation} s_3=\mbox{dim Im }P_q,\ \lambda _3=\mbox{dim Im }P_1 .
\label{eq:eqdim3}
\end{equation}

We use the result obtained by Sudbery.
For  $x \in \Pi _{\Gamma}$ put $y =q^{-1}R_{13}x$ then $y\in \Pi _{\L}$
and
\begin{equation} \begin{array}{lllllr}
R_{12}x & = & x & R_{23} x & = & qx+y\\
R_{12}y & = & -qx+qy & R_{23} y& = & y
\end{array}
\label{eq:sud}
\end{equation}
The first and last equations are obvious.
According to (\ref{eq:8}), $(R_{12}R_{23}-qR_{23}+q^2)P_{\Gamma}=0$, this
shows the third equation $R_{23}x=qx+y$, the second one follows from it.
 Analogously we have, for
$ x^{+}\in \Pi _{\Gamma}^{+}$ and $y^{+}=q^{-1}R_{13}^{*}x$,
then $y^+\in \Pi _{\L}^{+}$ and
\begin{equation} \begin{array}{lllllr}
R_{12}^{*}x^{+} & = & x^{+} & R_{23}^{*}x^{+} & = & qx^{+}+y^{+}\\
R_{12}^{*}y^{+} & = & -qx^{+}+qy^{+} & R_{23}^{*}y^{+} & = & y^{+}
\end{array}
\label{eq:sudac}
\end{equation}

Now  we consider  the operator
$\overline{R}=\theta (R^{*}\otimes R^{-1})\theta^{-1} $.
It this easy to check that
\bm
\begin{array}{lll}
\overline{R}_{12}&=&\theta (R_{12}^{*}\otimes R_{12}^{-1})\theta^{-1} \\
\overline{R}_{23}&=&\theta (R_{23}^{*}\otimes R_{23}^{-1})\theta^{-1}
\end{array}
\label{eq:rbar}
\em
 By the isomorphism $\theta :
(V^{*})^{\otimes 3}\otimes V^{\otimes 3}\rightarrow
(V^{*}\otimes V)^{\otimes 3}$
 we can regard $\overline{R}_{12}$ as
$R_{12}^{*}\otimes R_{12}^{-1}$ acting upon
$(V^{*})^{\otimes 3}\otimes V^{\otimes 3}.$ Consequently $\R$ is a Yang-Baxter
operator.

By considering the eigenvectors of $R$ which form a basis on $V\ot V$
we get  $(\overline{R}-1)(\overline{R}-q)(\overline{R}-q^{-1})=0$
 and $\mbox{dim Im }(\overline{R}-1)=2\lambda _{2}s_{2}$.
Since $\lambda _{2}+s_{2}=n^{2}$, by using (\ref{eq:hai}) we have
\begin{equation}
e_{2}=\mbox{dim }E_{2}=n^{4}-\mbox{dim Im }(\overline{R}-1)=\lambda _{2}^{2}
+s_{2}^{2} .
\label{eq:dim2}
\end{equation}
\begin{lem}
\label{sec-mainlem}
 Let $\Phi =\overline{R}_{12}\overline{R}_{23}\overline{R}_{12}
- - - -\overline{R}_{12}\overline{R}_{23}-\overline{R}_{23}\overline{R}_{12}
+\overline{R}_{12}+\overline{R}_{23}-1$, then
\begin{equation}
\mbox{\rm Im}\Phi =\mbox{\rm Im }(\overline{R}_{12}-1)\cap
 \mbox{\rm Im }(\overline{R}_{23}-1).
\label{eq:intersec}
\end{equation}
\end{lem}
 {\bf Proof.}
 Since $\overline{R}$ is a Yang-Baxter operator
$$\Phi =(\overline{R}_{12}-1)(\overline{R}_{23}\overline{R}_{12}-
\overline{R}_{23}+1)=(\overline{R}_{23}-1)(\overline{R}_{12}\overline{R}_{23}-
\overline{R}_{12}+1).$$ Hence $\mbox{Im }\Phi
\subset (\mbox{Im }(\overline{R}_{12}-1)\cap
\mbox{Im }(\overline{R}_{23}-1))=W.$

 If we can show that $W\cap \mbox{Ker }\Phi ={0},$ then $\mbox{dim }W+
\mbox{dim Ker }\Phi
\leq n^{6} $(since they are subspaces of $(V^{*})^{\otimes 3}\otimes
V^{\otimes 3})$.
But $\mbox{dim Im }\Phi +\mbox{dim Ker }\Phi =n^{6}$, therefore $\mbox{dim }W=
\mbox{dim Im }\Phi $, which means $W=\mbox{Im }\Phi$.

Assume $L:=W\cap \mbox{Ker}\Phi \not= \{ 0\} $. Denote
$\overline{R}_{13}=\overline{R}_{12}\overline{R}_{23}\overline{R}_{12}=
\overline{R}_{23}\overline{R}_{12}\overline{R}_{23}$, then
$$\Phi \overline{R}_{13}=\overline{R}_{13}\Phi $$
since
$\overline{R}_{13}\overline{R}_{12}=\overline{R}_{23}\overline{R}_{13}$,
 $\overline{R}_{12}\overline{R}_{13}=\overline{R}_{13}\overline{R}_{23}$.
Hence L is invariant under  $\overline{R}_{13}$,
therefore there exists a vector
$x\not= 0$ in L so that $\overline{R}_{13}x=\lambda x$.

 If $\lambda = -1$ then
\begin{eqnarray*}\begin{array}{rl}
\Phi x=&-2x-\overline{R}_{12}\overline{R}_{23}x
- - -
-\overline{R}_{23}\overline{R}_{12}x+\overline{R}_{12}x+\overline{R}_{23}x
\\
=& -2x+\overline{R}_{12}^{-1}x+\overline{R}_{23}^{-1}x+\overline{R}_{12}x
+\overline{R}_{23}x .\end{array}\end{eqnarray*}
Since $x\in W =
\mbox{Im}(\overline{R}_{12}-1)\cap\mbox{Im}(\overline{R}_{23}-1)$,
we get $(\R_{12}x-q^{-1})(\R_{12}-q)=0$ and
$(\R_{12}^2-(q+q^{-1})\R_{12}+1)x=0$. Thus
\begin{equation}\begin{array}{l}
(\overline{R}_{12}^{-1}+\overline{R}_{12}-q-q^{-1})x=0\\
(\overline{R}_{23}^{-1}+\overline{R}_{23}-q-q^{-1})x=0\end{array}
\label{eq:hai1}
\end{equation}
Hence $\Phi x=2(q+q^{-1}-1)x=0,\ q+q^{-1}-1\not= 0,$ therefore $x=0$, which is
a contradiction.

Let $\lambda \not= -1$. Since $\R_{13}=\R_{23}\R_{12}\R_{23}x=\lambda x$
we get
\begin{equation}(\lambda \overline{R}_{23}^{-1}
- - -
-\overline{R}_{23})x=(\overline{R}_{12}\overline{R}_{23}-\overline{R}_{23}
)x
=(\overline{R}_{12}-1)\overline{R}_{23}x
\in \mbox{Im}(\overline{R}_{12}-1)\label{eq:ref}\end{equation}
According to (\ref{eq:hai1}),
$(\lambda \overline{R}_{23}^{-1}-\overline{R}_{23})x=\lambda (q+q^{-1})x-
(\lambda +1)\overline{R}_{23}x$.
And since $x\in W\subset\mbox{Im}(\overline{R}_{12}-1)$ we get
$\overline{R}_{23}x\in \mbox{Im}(\overline{R}_{12}-1)$. Therefore
we get $$z:=(\overline{R}_{23}\overline{R}_{12}-\overline{R}_{23}+1)x=
(\lambda\overline{R}_{12}^{-1}-\overline{R}_{23}+1)x \in
\mbox{Im}(\overline{R}_{12}-1).$$
Then $(\overline{R}_{12}-q)(\overline{R}_{12}-q^{-1})z=0$ and
  $(\R_{12}-1)z=(\overline{R}_{12}-1)(\overline{R}_{23}
\overline{R}_{12}-\overline{R}_{23}+1)x=\Phi x=0$
hence $z=0$, that means
\begin{equation}
(\overline{R}_{23}\overline{R}_{12}-\overline{R}_{23}+1)x=0.
\label{eq:a1}
\end{equation}
 Analogously we have
\begin{equation}
(\overline{R}_{12}\overline{R}_{23}-\overline{R}_{12}+1)x=0.
\label{eq:a2}
\end{equation}
 Multiplying (\ref{eq:a1}) by $\overline{R}_{12}$ from the left and adding
(\ref{eq:a2}) we get $\overline{R}_{13}x=
\overline{R}_{12}\overline{R}_{23}\overline{R}_{12}x
 =-x$ which is a contradiction. Thus we must have  $L=\{ 0\},$ which means
$\mbox{Im}\Phi =W$. \hfill Q.E.D.

\begin{cor}
 \label{sec-cor37}
\begin{equation}
\mbox{\rm Im }\Phi \oplus \mbox{\rm Ker }
\Phi =(V^{*})^{\otimes 3}\otimes V^{\otimes 3}.
\end{equation}
\end{cor}
 Indeed $\mbox{Im }\Phi \cap \mbox{Ker }\Phi =\{ 0\}. $

\begin{pro}
\label{sec-hai}
 Let $\Pi =\Pi _{\Gamma}\oplus \Pi _{\L}$ , $\Pi ^{+}=\Pi _{\Gamma}^{+}\oplus
\Pi _{\L}^{+}$ then

$i)\Pi _{1}^{+}\otimes \Pi _{q},~ \Pi _{q}^{+}\otimes \Pi_{1}$ are subspaces
of $\mbox{Im }\Phi$.

$ii)\Pi _{\alpha}^{+}\otimes \Pi _{\alpha},~
\Pi _{\alpha}^{+}\otimes \Pi ,~ \Pi ^{+}\otimes \Pi _{\alpha},\ (\alpha =1,q)$
are subspaces of $\mbox{Ker }\Phi.$

$iii)$The subspace $\Pi ^{+}\otimes \Pi $ is invariant under the action
of $\Phi $
and decomposes into a direct sum of 4-dimensional $\Phi$-invariant subspaces.
In each of them the operator $\Phi$ has rank 1.
\end{pro}
{\bf Proof.}

$i)$For $x\otimes y \in \Pi _{1}^{+}\otimes \Pi _{q}, \overline{R}_{12}
x\otimes y =\overline{R}_{23}x\otimes y =q^{-1}x\otimes y$, hence
$\Phi x\otimes y =(q^{-1}-1)(q^{-2}-q^{-1}+1)x\otimes y,$ therefore $x\otimes y
\in \mbox{Im }\Phi$.

$ii)$ If $x\otimes y \in \Pi _{\alpha}^{+}\otimes \Pi _{\alpha},
\ (\alpha =1,q)$ then
$\overline{R}_{12}x\otimes y =\overline{R}_{23}x\otimes y =x\otimes y$
so that $\Phi x\otimes y =0$.
For $x\otimes y \in \Pi _{1}^{+}\otimes \Pi , R_{12}^{*}x=R_{23}^{*}x=qx$
and $P_{1}(qR^{-1})y=-R_{13}^{-1}P_qy=0$ hence
$$\Phi x\otimes y=x\otimes P_{1}(qR^{-1})y=0.$$

$iii)$ According to (\ref{eq:sud}) and (\ref{eq:sudac}) if $x_{j}$ form a basis
in
$\Pi _{\Gamma}$, $x_{i}^{+}$ form a basis in $\Pi _{\Gamma}^{+}$
then $y_{j}=q^{-1}R_{13}x_{j}$ form a basis in $\Pi _{\L},$
$y_{i}^{+}=q^{-1}R_{13}x_{i}^{+}$ form a basis in $\Pi _{\L}^{+},$
and the subspaces of $\Pi ^{+}\otimes \Pi $ spanned by the four vectors
$x_{i}^{+}\otimes x_{j}, x_{i}^{+}\otimes y_{j},
y_{i}^{+}\otimes x_{j}, y_{i}^{+}\otimes y_{j},$ are $\Phi $-invariant.
This shows the first part of $iii)$. The second one can be proved by direct
computation.\hfill Q.E.D.
\begin{dl}
\label{sec-dim}
If the operator R satisfies the Yang-Baxter equation
(\ref{eq:YB}), with $q$ satisfying condition (\ref{eq:cond}), then we have
\begin{equation}
e_{3}=n^{6}-4n^{2}s_{2}\lambda _{2}+2s_{3}\lambda _{3}+
(n^{3}-s_{3}-\lambda _{3})^{2}/4
\label{eq:dim3}
\end{equation}
\end{dl}
{\bf Proof.} From (\ref{eq:hai}), the relations on $E_{3}$ are
$\mbox{Im }(\overline{R}_{12}-1)+
\mbox{Im }(\overline{R}_{23}-1)$, hence
 $\mbox{dim }E_{3}= n^{6}-\mbox{dim }(\mbox{Im }(\overline{R}_{12}-1)+
\mbox{Im }(\overline{R}_{23}
- - - -1))= n^{6}-\mbox{dim Im }(\overline{R}_{12}-1)-\mbox{dim Im }
(\overline{R}_{23}-1)+
\mbox{dim }(\mbox{Im }(\overline{R}_{12}-1)\cap \mbox{Im }
(\overline{R}_{23}-1))= n^{6}-4n^{2}s_{2}
\lambda _{2}+\mbox{dim Im }\Phi $ (according to  (\ref{eq10})
and Lemma (\ref{sec-mainlem})).
  Since $\Pi \oplus \Pi _{1} \oplus \Pi _{q}=V^{\otimes 3}$, we have
\begin{equation}
\bigoplus_{\alpha ,\beta}
 (\Pi _{\alpha}^{+}\otimes \Pi _{\beta})=(V^{*})^{\otimes 3}\otimes
V^{\otimes 3},\  (\alpha ,\beta =\emptyset ,1,q)
\label{eq:eq251}
\end{equation}
 According to Corollary (\ref{sec-cor37}) and Proposition (\ref{sec-hai})
\begin{equation}
\mbox{dim } \mbox{Im }\Phi =\mbox{dim }
(\Pi _{1}^{+}\otimes \Pi _{q})+\mbox{dim }
(\Pi _{q}^{+}\otimes \Pi _{1})
+\mbox{dim }(\Pi ^{+}\otimes \Pi )/4
\label{eq:eq252}
\end{equation}
 Indeed, (\ref{eq:eq251}) is a decompositon of $(V^{*})^{\otimes 3}\otimes
V^{\otimes 3}$,
according to $iii)$ in Proposition (\ref{sec-hai}),
$$\Pi ^{+}\otimes \Pi =(\Pi ^{+}\otimes \Pi \cap \mbox{Im}\Phi )\oplus
(\Pi ^{+}\otimes \Pi \cap \mbox{Ker}\Phi )$$ is a
decomposition of $\Pi ^{+}\otimes \Pi .$
And $\mbox{dim}(\Pi ^{+}\otimes \Pi \cap \mbox{Im}\Phi )=\mbox{dim}(\Pi ^{+}
\otimes \Pi )/4,$
 which implies (\ref{eq:eq252}).
 From  Lemma (\ref{sec-sudbery}) and (\ref{eq:eqdim3}) one follows
$$\mbox{dim}\Pi _{1}=\lambda _{3}, \mbox{dim}\Pi _{q}=s_{3},
\mbox{dim}\Pi =n^{3}-s_{3}-\lambda _{3}.$$
Thus $\mbox{dim Im}\Phi= 2\lambda _{3}s_{3}+(n^{3}-s_{3}-\lambda _{3})^2/4$
which gives (\ref{eq:dim3}).\hfill Q.E.D.

  Theorem~\ref{sec-dim} shows that if the construction is Yang-Baxter then the
dimensions of $E_{2}$ and $E_{3}$ do not depend on the choice
 of the matrix $R$
and depend only on $\lambda _{i}$ and $s_{i}$. If the same independence would
take place for all the higher components, we could directly
calculate the Poincar\'{e} series of $E$ in many cases, avoiding the use of
the diamond lemma, which is not always applicable.

In the next section we investigate the polynomiality of some deformations
of $GL(n)$ however by using the diamond lemma.
 The following Corollary is useful.

  Let $E_{1}$ be spanned by $\{ z_{i}^{j}\}$,
  define an ordering of $z_{i}^{j}$ by
$z_{i}^{j}<z_{k}^{l}$ if either $i<k$ or $i=k,j<l.$
 Call a monomial z in $z_{i}^{j}$ normally ordered if for any
$z$', $ z"\in \{z_i^j\}$
in this monomial, $z$'$<z"$ iff $ z$' is to the left of $z"$.
 We consider the lexicographic order in the
polynomials in $z_{i}^{j}$, which corresponds to this order.
\begin{cor}
\label{sec-compatible}
 If the dimensions of the second and third
homogeneous components of S and $\Lambd$ coincide with the ones of S(V) and
$\Lambd$(V) respectively, then the dimension of the 2nd and 3rd homogeneous
components of E coincide with the ones of a polynomial ring of $n^{2}$
parameters. Moreover, if every monomial, which is not normally ordered,
 can be represented
as a sum of normally ordered monomials of strictly smaller order (in the
lexicographic order), then the normally ordered monomials form a basis of $E$.
\end{cor}
{\bf Proof.} If
$$s_{2}=n(n+1)/2,  \lambda _{2}=n(n-1)/2$$
$$s_{3}=n(n+1)(n+2)/6, \lambda _{3}=n(n-1)(n-2)/6$$
then, by using (\ref{eq:dim2}),(\ref{eq:dim3}), we have
\begin{equation}
\begin{array}{lllll}
e_{2}&=&\mbox{dim}E_{2}&=&n^{2}(n^{2}+1)/2\\
e_{3}&=&\mbox{dim}E_{3}&=&n^{2}(n^{2}+1)(n^{2}+2)/6
\end{array}
\label{eq:dime}
\end{equation}
 If the last condition ot the Corollary is verified then $z_{i}^{j}
z_{k}^{l},z_{i}^{j}\leq z_{k}^{l}$ generate $E_{2}.$ According
to (\ref{eq:dime}) they must be independent and so form the basis on $E_{2}.$
The reordering procedure of monomials of higher power is possible but
not unique.
Manin has shown [M1] that the different possible procedures will lead to the
same result if they do so for cubic monomials, in other words if the cubic
normally ordered monomials are independent. But they are so, according
to (\ref{eq:dime}). So that we can apply the diamond lemma for the algebra $E$.
\hfill Q.E.D.

{\bf Remark.} In general the normally
ordered monomials  will not span $E$, even if  the subspace $E_{2}$
is spanned by normally ordered monomials. If they span $E_{2}$  they may
 even not span  $E_{3}.$
 An example is the following quadratic algebra, $A=k<x,y>/
yx=x^2+y^2+xy$. The normally ordered monomials span $A_2$ but do not span
$A_3$. Note that, however its Poincar\'{e} series are correct [EOW], p. 1.
The next section considers this question for the deformations of
$GL(n)$.

\section{Polynomiality of some deformations}
 There are mainly three series of deformation of $GL(n)$. Artin, Schelters and
Tate [ATS] found the multiparameter
deformation, which include  Sudbery's and
Manin's deformations [S], [DMMZ]. Cremer and Gervais[CG] found a
deformation, which is
associated with the non-linearly extended Virasoro algebras. The Jordan
 deformation of $GL(2)$
was found by Manin[DMMZ] and was systematically studied by Wess and
others [EOW]. In
this section we investigate the polynomiality of the deformation
obtained by Creme and Gervais.
 At the end of the section we form a new deformation for $GL(3)$.

\subsection{The deformation of E.Cremer and J.-L.Gervais.}
 This deformation is given by the matrix
\begin{eqnarray}
R=&\sum_{k,i>j}(e_{kk}^{kk}+q^{n+j-i}e_{ij}^{ji}+q^{i-j}e_{ji}^{ij}+
(1-q^{n})e_{ji}^{ji}+ \nonumber \\
&\sum_{r=1}^{i-j-1}
((q^{n}-1)q^{-r}e_{ji}^{j-r,i+r}+ (1-q^{n})q^{r}e_{ij}^{i-r,j+r}))
\label{eq:CG}
\end{eqnarray}
Note that $R=P\cdot\rho$, where $\rho$ is the operator defined in [CG], p.625,
$P$ is the usual twist operator.
$R$ satisfies the Yang-Baxter equation and the equation $$
(R-1)(R+q^{n})=0.$$ We assume that
$q\not= \sqrt{1}.$ The subspace relations $R_{\mbox{s}}$ and $R_{\Lambd}$
can be spanned by the vectors
\begin{equation}
u_{ij}=x_{i}x_{j}-q^{i-j-n}x_{j}x_{i}-(q^{n}-q^{-n})q^{i-j-r-rn}x_{i+r}x_{i-r}
+p(1-q^{n})x_{\frac{i+j}{2}}x_{\frac{i+j}{2}}
\label{eq:eq43}
\end{equation}
$$(1\leq i<j\leq n)$$

where
\[ p=\left\{ \begin{array}{llll}
0&\mbox{if}&i+j& \mbox{odd}\\
\frac{i-j}{2}+n-\frac{n(i-j)}{2} &\mbox{if}& i+j& \mbox{even}
\end{array}\right. \]
and
\begin{equation}\left\{
\begin{array}{lll}
v_{ii}&=&x_{i}x_{i}\\
v_{ij}&=&x_{i}x_{j}+q^{i-j}x_{j}x_{i},\ (1\leq j < i\leq n)
\end{array}\right.
\label{eq:eq44}
\end{equation}
respectively. We use the following lemma for investigating the compatibility
of $S$ and $\Lambd$.
\begin{lem}
\label{sec-lemgurb}([G]p.556)
Let $P_{\mbox{s}}(t),P_{\Lambd}(t)$ be the Poincar\'{e} series of S and
$\Lambd$
respectively. If the  matrix R satisfies the Yang-Baxter
equation (\ref{eq:YB}), then
\begin{equation}
P_{\mbox{s}}(t)P_{\Lambd}(-t)=1
\label{eq:gura}
\end{equation}
\end{lem}
\begin{dl}  The quantum matrix bialgebra E determined by S and
$\Lambd ,$ described in (\ref{eq:eq43}) and (\ref{eq:eq44}), is compatible.
\end{dl}
{\bf Proof.} Denote the  Poincar\'{e} series of $\Lambd $ and S by
$P_{\Lambd}(t)$ and $P_{\mbox{s}}(t)$ respectively. According to lemma
(\ref{sec-lemgurb})
we have $P_{\Lambd}(t)P_{\mbox{s}}(-t)=1.$ From (\ref{eq:eq44}) we have
$P_{\Lambd}
=(1+t)^{n}.$ Hence $P_{\mbox{s}}(t)=\frac{1}{(1-t)^{n}}$ so that $\Lambd $
and S
are compatible.\par
As we have seen in section 2, the subspace relation on E are $R_{\mbox{s}}^{*}
\otimes R_{\mbox{s}}\oplus R_{\Lambd}^{*}\otimes R_{\Lambd}$ when
$E_{1}\otimes E_{1}$ and $V^{*}\otimes V^{*}\otimes V\otimes V$ are
identified.\par
 Obviously the subspace $R_{\Lambd}^{*}$ is spanned by
\begin{equation}
v^{ij}=\xi ^{i}\xi ^{j}-q^{n-i+j}\xi ^{j}\xi ^{i},\ 1\leq j < i\leq n.
\label{eq:eq45}
\end{equation}
 We show that $R_{\mbox{s}}^{*}$ can be spanned by vectors of the type
\begin{equation}
u^{ij}=\xi ^{i}\xi ^{j}+q^{n-i+j}\xi ^{j}\xi ^{i}+\sum_{1\leq r\leq
j-1,r\leq n-i}c_{r}^{ij}\xi ^{j-r}\xi ^{i+r}
\label{eq:eq46}
\end{equation}
$(1\leq j\leq i\leq n)$.\\
 In fact, we must find coefficents $c_{r}^{ij}$
so that $\langle u^{ij}|u_{kl}\rangle =\delta _{k}^{i}\delta _{l}^{j}
(i\geq j,k<l)$. Fix i and j then for every composition of $c_{r}^{ij}$,
$\langle u^{ij}|u_{kl}\rangle =0$ if either $k+l\not= i+j$ or $l>j.$
Hence we have to find $c_{r}^{ij} $ obeying only the following system
of equations
\begin{equation}
\langle u^{ij}|u_{kl}\rangle =0\ (s<n-i+1, s<j)
\label{eq:eq47}
\end{equation}
whose rank is min(j-1,n-i).\par
 Thus we get linear equations for $c_{r}^{ij}$, where the number of parameters
equals the number of equation, and what is more, the coefficents $a_{s}^{r}$
of $c_{r}^{ij}$ in the s-th equation is equal 1 if $r=s$, and 0 if $r<s$. Hence
(\ref{eq:eq46}), (\ref{eq:eq47}) has a unique solution .

 Using (\ref{eq:eq43}-\ref{eq:eq46}), we get the following relations in E:
\begin{equation}
z_{i}^{l}z_{j}^{k}+\sum c_{r}^{'kl} z_{i}^{l-r}z_{j}^{k+r}+\sum_{m<n,m<i}
p_{uv}^{'mn} z_{m}^{u}z_{n}^{v}=0, (1\leq j<i\leq n, 1\leq l\leq k \leq n)
\label{eq:eq48}
\end{equation}

\begin{equation}
z_{i}^{k}z_{j}^{l}+\sum c_{r}^{''kl} z_{i}^{l-r}z_{j}^{k+r}+\sum_{m<n, m<i}
p_{uv}^{''mn} z_{m}^{u}z_{n}^{v}=0, (1\leq j<i\leq n, 1\leq l\leq k \leq n)
\label{eq:eq49}
\end{equation}

\begin{equation}
z_{i}^{k}z_{i}^{l}-q^{kl}z_{i}^{l}z_{i}^{k},(1\leq i\leq n,\ 1\leq l<k\leq n)
\label{eq:eq50}
\end{equation}
 By induction (\ref{eq:eq48}), (\ref{eq:eq49}) can be written in the following
form
\begin{equation}
z_{i}^{l}z_{j}^{k}+\sum_{m<n,m<i}\overline{p}_{uv}^{mn} z_{m}^{u}z_{n}^{v}=0,
(1\leq j<i\leq n)
\label{eq:eq51}
\end{equation}
 And so (\ref{eq:eq50}), (\ref{eq:eq51}) are generating relations for E.
According
to the diamond lemma normally ordered monomials span E. Finally, using
corollary (\ref{sec-compatible}), one finds that E is compatible.
\hfill Q.E.D.
\subsection{The Jordan deformation of $GL(3)$.}
\label{sec-jord}
In [DMMZ] Manin formed the non-standard deformations of $GL(2)$. One of them
is the Jordan deformation, which was systematicaly studied in [EOW].
Here we give its analogue for $GL(3)$. We show that the new bialgebra is still
compatible.
 Let
\begin{equation}
S=k<x_{1},x_{2},x_{3}>\left/ \left(
\begin{array}{r} x_{2}x_{1}-x_{1}x_{2}+p_{12}x_{2}x_{2}=0\\
x_{3}x_{2}-x_{2}x_{3}=0\\
x_{3}x_{1}-x_{1}x_{3}+p_{13}x_{2}x_{2}=0
\end{array}\right) \right.
\label{eq:eq52}
\end{equation}
\begin{equation}
\Lambd ^{!}=k<\xi ^{1},\xi ^{2},\xi ^{3}>\left/ \left(
\begin{array}{r}
\xi ^{3}\xi ^{1}-\xi ^{1}\xi ^{3}+q_{31}\xi ^{1}\xi ^{1}=0\\
\xi ^{3}\xi ^{2}-\xi ^{2}\xi ^{3}+q_{32}\xi ^{1}\xi ^{1}=0\\
\xi ^{2}\xi ^{1}-\xi ^{1}\xi ^{2}=0
\end{array}\right) \right.
\label{eq:eq53}
\end{equation}
where  $\Lambd ^{!} $ is the dual of $\Lambd $,(see [M]).\\
Let $E=E(S,\Lambd )$, then the matrix R is given by
$(q_{ij}=-q_{ji},p_{ij}=-p_{ji})$
\begin{equation}R=\left|
\begin{array}{lllllllll}
1 &. &. &. &. &. &. &. &. \\
. &. &. &1 &. &. &. &. &.\\
q_{13} &. &. &. &. &. &1 &. &.\\
. &1 &. &. &. &. &. &. &.\\
. &. &. &. &1 &. &. &. &.\\
q_{23} &. &. &. &. &. &. &1 &.\\
q_{31} &. &1 &. &. &. &. &. &.\\
q_{32} &. &. &. &. &1 &. &. &.\\
p_{13}q_{31} &p_{12} &p_{13} &p_{21} &. &. &p_{31} &. &1
\end{array}\right|
\end{equation}
is Yang-Baxter and $R^{2}=1$,
it is easy to show that S and $\Lambd $ are compatible.

 Let us define the ordering in E as follows, $z_{i}^{j}<z_{k}^{l}$ iff
$i>k$ or $i=k,j>l$. Then we have
\begin{dl}  E is compatible and the normally ordered monomials with
respect to the given order form a basis of E.
\end{dl}
{\bf Proof.} A direct computation shows that normally ordered monomials
span E. By applying (\ref{sec-compatible}) we get the result.\hfill Q.E.D.

\begin{center} {\bf ACKNOWNLEDGEMENT.} \end{center}

 The autor likes to thank Professor Yu.Manin for pointing him to these
problems, Professor J.Wess, who makes it possible that this work was written
in his institute and Professor B.Pareigis, who gave him very many useful
advices.
\end{document}